# Irreversible Aging Dynamics and Generic Phase Behavior of Aqueous Suspensions of Laponite

A. Shahin and Yogesh M Joshi*

Department of Chemical Engineering, Indian Institute of Technology Kanpur, Kanpur 208016. INDIA.

* Corresponding author, E-Mail: joshi@iitk.ac.in

Abstract

In this work we study the aging behavior of aqueous suspension of Laponite having 2.8 weight % concentration using rheological tools. At various salt concentration all the samples demonstrate orientational order when observed using crossed polarizers. In rheological experiments we observe inherent irreversibility in the aging dynamics which forces the system not to rejuvenate to the same state in the shear melting experiment carried out at a later date since preparation. The extensive rheological experiments carried out as a function of time elapsed since preparation demonstrate the self similar trend in the aging behavior irrespective of the concentration of salt. We observe that the exploration of the low energy states as a function of aging time is only kinetically affected by the presence of salt. We estimate that the energy barrier to attain the low energy states decreases linearly with increase in the concentration of salt. The observed superposition of all the elapsed time and the salt concentration dependent data suggests that the aging that occurs in low salt concentration systems over a very long period is qualitatively similar to the aging behavior observed in systems with high salt concentration over a shorter period.

**I. Introduction:**

In soft glassy (non-ergodic) materials, microstructure evolves to occupy low energy sections of the energy landscape in its search for a possible thermodynamic equilibrium.[1] The time required to attain a particular micro-structural state, however depends on the initial state and the nature of the path that connects these two states through an energy landscape. Greater degrees of freedom associated with the primary arrested entity (particle) of the glassy materials, such as nature of charges and charge distribution on the particle along with its shape and size distribution, may lead to a very intricate energy landscape. Consequently, the corresponding glassy material may show a very rich micro-structural phase behavior. Aqueous suspensions of smectite clays are known to undergo ergodicity breaking and to form soft pasty materials that have significant applications in chemical, petroleum, cosmetic/pharmaceutical, food, etc. industries.[2-5] Particularly aqueous suspension of Laponite is known to show a very rich phase behavior as a function of its concentration as well as the concentration of salt, and is a subject of topical interest.[6-33] However, the research work of more than a decade is divided over the possible microstates and their nomenclatures. Such contradiction is principally due to complexities involved in the charge distribution over the anisotropic particle of Laponite. In this article we have carried out extensive rheological study of aqueous suspension of Laponite having varying concentration of salt to investigate the effect of time elapsed since preparation on the phase behavior.

Laponite is a smectite hectorite clay and belongs to a structural family known as 2:1 phyllosilicates.[2] The chemical formula of Laponite is $Na_{+0.7}[(Si_8Mg_{5.5}Li_{0.3})O_{20}(OH)_4]_{-0.7}$.[34] In a unit cell of the Laponite crystal, six octahedral magnesium ions are sandwiched between two layers of four tetrahedral silicon atoms. These groups are balanced by twenty oxygen atoms and four hydroxyl groups.[34] Isomorphic substitution of divalent magnesium by monovalent lithium induces deficiency of positive charge within the layer that gives a permanent negative charge to the surface of Laponite.[2] Laponite particle has a disc like shape with diameter of 25-30 nm and thickness of 1 nm.[35] In the powder form, Laponite particles are present in stacks with sodium ions in the interlayer gallery that balance the negative surface charge. In an aqueous medium, sodium ions dissociate rendering a net negative charge to the Laponite surface. The edge (rim) of the disc is composed of hydrous oxides such as Mg-OH and Si-OH. In the low pH medium protonation of the edge is preferred rendering a $-OH_2^+$ (positive) charge to the edge. Intensity of the positive charge on the edge goes on decreasing with the increase in pH. In the limit of strong basic environment deprotonation dominates wherein $H^+$ ions dissociate giving a $-O^-$ (negative) charge to the edge.[2, 36] According to Tombacz and Szekeres[37] the point of zero charge for Si-OH group is associated with pH between 4.5 and 5.5., while according to Kosmulski,[38] for MgOH the iso-electric point is expected to be above pH of 10. In a Laponite particle, the intermediate octahedral layer contains Mg (and Li in smaller amount), which forms complex with OH. In the outer tetrahedral layer Si has stronger association with oxygen on the surface of the Laponite. Therefore edge of the Laponite particle can be considered to have predominantly MgOH groups. As reported by Martin and coworkers,[39] Laponite technical bulletin[40] claims that the edge of Laponite particle contains predominantly MgOH and is positive below pH of 11. Tawari and coworkers[36] estimated the edge charge using pH and conductivity measurements and found the same to be positive having magnitude 50e2 per Laponite particle at pH of 9.97. To avoid leaching of magnesium from the Laponite particle in an acidic environment,[23, 41] pH of the suspension in an aqueous media is generally kept around 10. In absence of any externally added salt, there is a repulsion among the Laponite particles, which is believed to be causing ergodicity breaking at a very low concentration of salt.[13, 20, 42, 43] However, various experimental and simulation results suggest that the possibility of the attractive interactions between the edge and the surface can not be ruled out.[18, 36, 44] Addition of external salt such as NaCl increases the concentration of cations, which screen the negative charge on the surface and reduce the repulsion among the Laponite particles. Overall, the anisotropic disc-like shape of the Laponite particle and the uneven charge distribution on the same along with effects arising from the concentration of Laponite and that of the externally added salt lead to a rich spectrum of microstructures, which evolve with time in a very complicated energy landscape.[14, 25, 45, 46]

In the literature, various groups have proposed different versions of the phase diagrams as a function of concentration of Laponite and that of the salt (NaCl).[4, 5, 15, 18, 20, 27, 42, 47] It is generally observed that soon after mixing around 1 vol. % or more Laponite in water, a soft solid state is formed that withstands its own weight. The microstructure of Laponite suspension is known to be very sensitive to the sample preparation protocol.[12, 18] At very low ionic concentration, ergodicity breaking is proposed to be caused by dominating repulsive interactions among the Laponite particles due to overlapping of double layers,[20, 22, 30, 42, 48, 49] and the corresponding state is identified as a repulsive glass. Tanaka *et al.* [42] proposed that with increase in the concentration of salt, reduced Coulomb repulsion among the particles leads to formation of an attractive glass. With further increase in the concentration of salt, van der Waals attraction prevails and the system enters a gel state.[42] Some groups proposed that the ergodicity breaking in aqueous Laponite suspension is due to positive edge – negative surface attractive interactions which leads to the gel formation.[18, 44] Cocard and coworkers[47] studied evolution of elastic and viscous modulus for Laponite suspensions having varying concentration salt. They observed that the evolution of both the moduli shifts to lower ages for the systems having higher concentration of salt suggesting that concentration of salt affects the rate of gelation. Ruzicka *et al.* proposed that at low concentration of salt and at low concentrations of Laponite, the system forms an inhomogeneous structure which they represent as gel, while at low concentrations of salt and high concentrations of Laponite they observed homogeneous structure, which they represent as attractive glass.[50] Li *et al.*[8] suggested that the Laponite platelets interact via an attractive potential on short distances but repulsive on longer distances emphasizing the role of short-range attraction for the process of aggregation at higher volume fractions of Laponite. Recently Jabbari-Farouji and coworkers [15, 45] proposed that the repulsive glassy state obtained in the Laponite suspension having high Laponite concentration and no added salt is similar to that observed in hard sphere glasses. For the moderate concentrations of Laponite, they proposed that the gel state and the glass state are part of the same energy landscape and the system can get arrested in the either state.[15] Overall, there are multiple proposals and over 15 years of research in this area has not lead to much consensus to unify various observations.

In this paper, we investigate evolution of elastic and viscous modulus of the aqueous Laponite suspensions having different concentrations of salt as a function of aging time at regular intervals for more than 18 days of idle time after the preparation of the sample. We observe that the suspensions undergo irreversible aging with subsequent evolution curves shifting to lower ages for the experiments carried out on a progressively later date since preparation. Experiments with various salt concentrations show self similar behavior leading to *salt concentration–idle time–aging time superposition.*

**II. Experimental Procedure:**

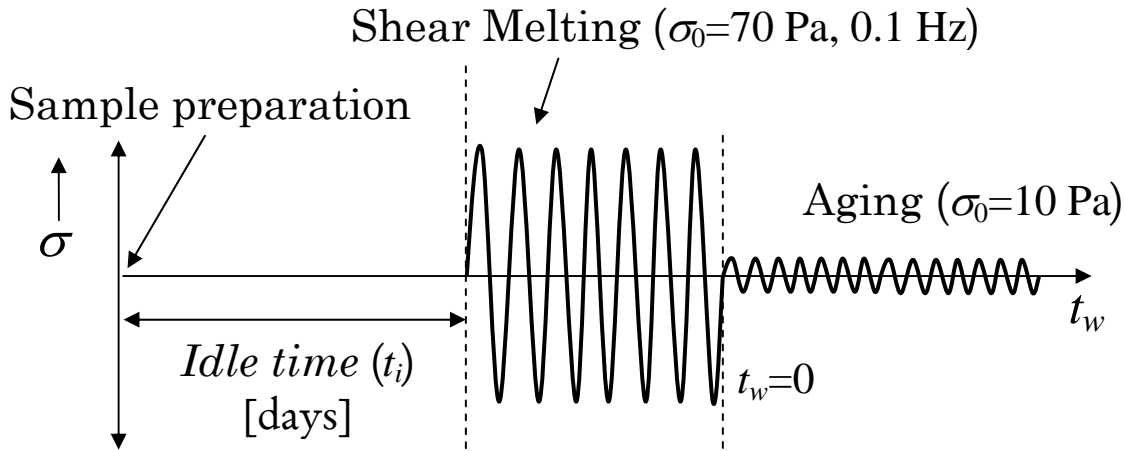

**Figure 1.** Experimental protocol followed in the present study. An ordinate represents shear stress axis while an abscissa represents time axis.

Laponite RD®, procured from Southern Clay Products Inc., was dried in the oven at 120°C for 4 hours. The desired amount clay was mixed with deionized water, which has pH 10 and predetermined amount of salt (NaCl), using an ultra turrex drive for a period of 45 minutes. The suspension was then stored in a sealed polypropylene bottle. In this work we have used 2.8 weight % of Laponite suspension having five concentrations of salt (NaCl) in the range 0.1 to 7 mM. The experimental procedure is schematically described in fig. 1. After preparing the suspension, we carried out the rheological experiments at the desired idle times ranging from 3 days up to minimum 18 days. We also checked for the chemical stability of the suspension by carrying out the complexometric titrations using EDTA with Eriochrome black T indicator.[23, 51] We did not find any measurable amount of $Mg^{2+}$ ions in the suspension demonstrating the same to be chemically stable.

In the rheological study, in each experiment a fresh sample at certain idle time was loaded in a couette cell using an injection syringe. The suspension partly melts in this step. After the thermal equilibrium was attained the sample was completely shear melted in an oscillatory test with shear stress amplitude of 70 Pa and a frequency of 0.1 Hz. The time at which shear melting was stopped marked the beginning of the aging experiment ($t_w$=0) as shown in fig. 1. In the subsequent aging experiments, oscillatory shear stress having magnitude 10 Pa and frequency 0.1 Hz was applied to the shear melted suspensions to record the evolution of their viscoelastic properties. It should be noted that aqueous suspension of Laponite continuously undergoes structural evolution increasing its viscosity and the elasticity causing broadening of the rheological linear response regime as a function of time. However, at very low age, since time required for the stress sweep experiment is usually more than the age of the sample, the suspension does not allow estimation of linear response regime. While the aging experiments were indeed in the linear response regime at higher ages, we confirmed that the oscillatory response of strain was always harmonic irrespective of the age. It was usually observed that the third harmonic contributed negligibly to the response.[46] In this work, the rheological experiments were carried out using a stress controlled rheometer AR–1000 (couette geometry with a bob diameter of 28 mm and a gap of 1 mm). The free surface of the suspension was covered with a thin layer of low viscosity silicon oil to avoid any contact of the same with air. We also measured the ionic conductivity of the suspensions using a CyberComm 6000 (Thermo Fisher Scientific Inc.) conductivity meter. All the experiments mentioned in this manuscript were carried out at 10°C.

**III. Results:**

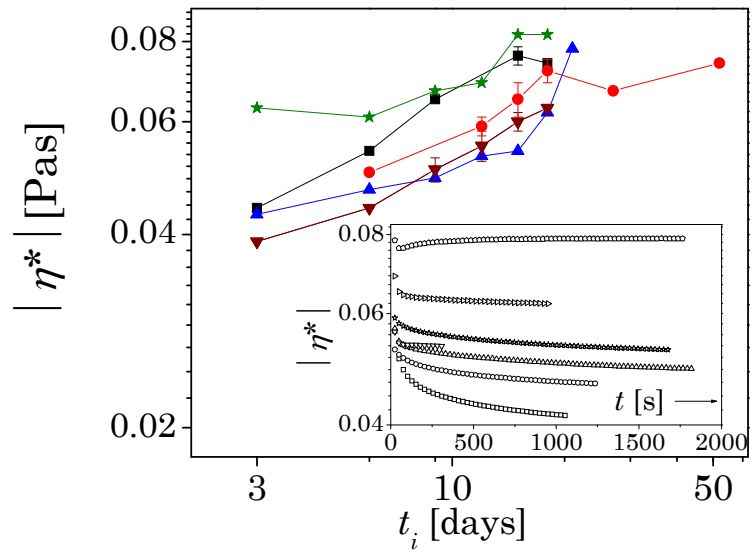

**Figure 2**: Plateau values of the complex viscosity in a shear melting step as a function of idle time ($t_i$) for various salt concentrations (Filled symbols; squares: 0.1mM; circles: 1mM; up triangles: 3mM; down triangles: 5mM; stars: 7mM). The inset shows day wise shear melting data for the 3 mM Laponite suspension (From Bottom to Top): $t_i$ = 3, 6, 9, 12, 15, 18, and 21 days. Rejuvenation was carried out at stress amplitude of 70 Pa.

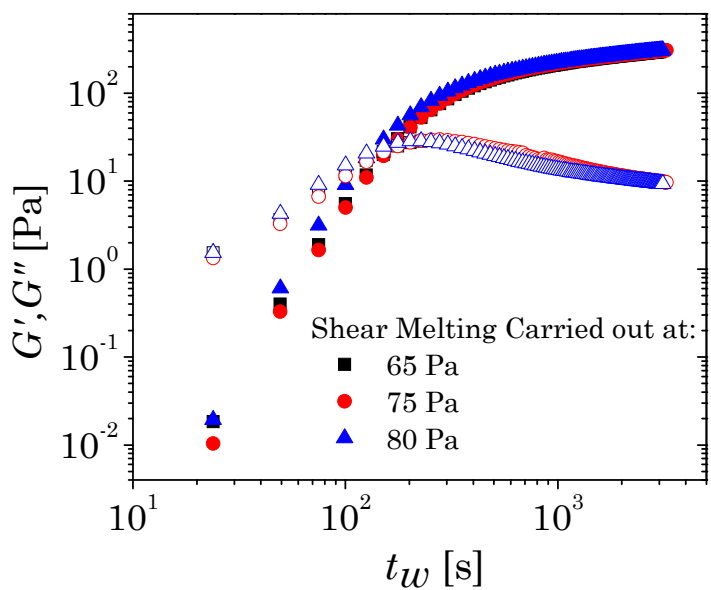

**Figure 3.** Evolution of Storage ($G'$: filled symbols) and Loss modulus ($G''$: open symbols) [Stress amplitude 10 Pa, frequency 0.1 Hz] for a 5 mM suspension on $t_i$ =3 days for different rejuvenating stress amplitudes applied during the shear melting.

As shown in fig. 1, in each experiment, the shear melting step preceded the aging step. Inset in fig. 2 describes a shear melting behavior for a Laponite suspension having 3 mM NaCl concentration as a function of idle time. It can be seen that complex viscosity reaches a plateau value in less than 10 min. It can be seen from an inset of figure 2, that the system reaches a higher value of plateau viscosity for the shear melting experiments carried out at greater idle times. Figure 2 shows variation of the plateau value of complex viscosity as a function of idle time. For all the explored concentrations of salt the plateau viscosity increases with increase in the idle time. This behavior demonstrates the partial irreversibility in the aging dynamics of aqueous suspension of Laponite, which causes complex viscosity in the shear melting step not to reach the same plateau value. It must be mentioned that, in the shear melting step we verified that the sample indeed underwent the *complete rejuvenation*. Such verification can be carried out by applying higher magnitude of shear stress amplitude to the sample in the shear melting step and studying the subsequent evolution of elastic and viscous modulus. Figure 3 shows evolution of elastic and viscous modulus after carrying out the shear melting at three different stresses for 5 mM Laponite suspension on $t_i$=3 days. We observed that in the shear melting step the plateau value of complex viscosity was lower for greater stress. It can be seen that irrespective of the value of the rejuvenation stress in the mentioned range, the subsequent evolution of both the moduli is not affected. This observation is in agreement with Joshi *et al.*,[14] who showed that evolution of the elastic modulus shows identical behavior irrespective of the amplitude of stress in the shear melting step when the shear melting is performed at the stress amplitudes greater than the yield stress of the material.

Subsequent to the shear melting experiments, the aging experiments were performed as shown in fig.1. Figures 4a and 4b show respective evolutions of $G'$ and $G''$ as a function of aging time ($t_w$), at various idle times ($t_i$) up to 52 days, for a system with 1 mM salt concentration. It can be seen that the evolutions of both $G'$ and $G''$ with respect to aging time gets systematically shifted to low aging times for the experiments carried out at a later date since preparation (higher idle time, $t_i$). Similar curvatures of all the aging curves result in a natural superposition by horizontal shifting as described by the superposing gray points on $t_i$=18 days data. We have also plotted the superposition of $G''$, which is shown in fig. 4b, on fig. 4a. It can be seen that $G''$ shows a two step evolution with an initial increase followed by a decrease with the aging time. Correspondingly, $G'$ also shows a two step evolution, wherein it undergoes a rapid rise in the first step, and after crossing over $G''$ around its maximum, it undergoes a slow power law increase in the second step. The crossover of $G'$ and $G''$ in an aging experiment is generally considered as an ergodicity breaking point,[52] when the system shows a transition from a liquid-like to a solid (glass)-like behavior. Interestingly, for the experiment carried out for idle time $t_i$=52 days, the system does not show the liquid regime ($G'<G''$) and directly enters the glassy regime ($G'>G''$).

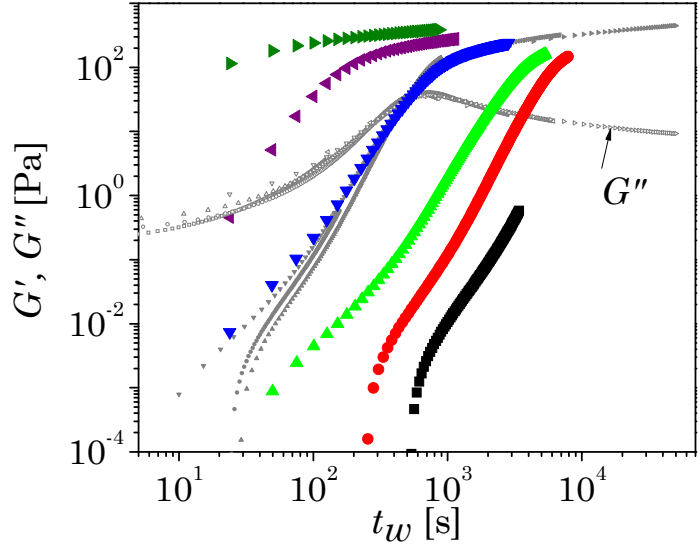

(a)

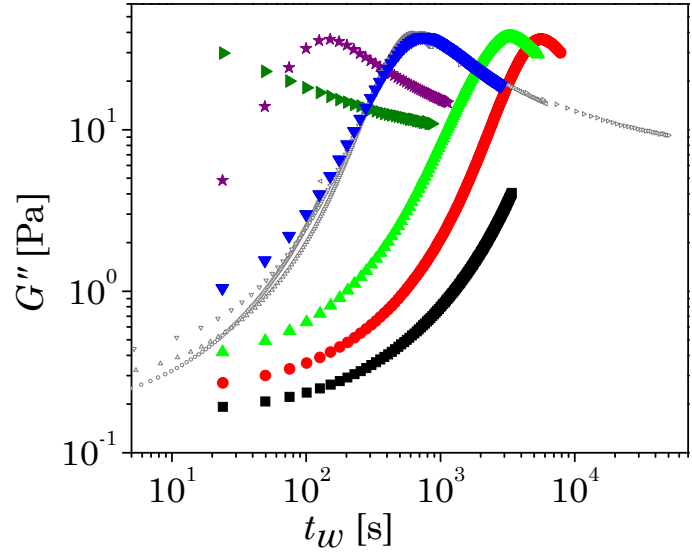

(b)

**Figure 4**: Evolution of (a) Storage and (b) Loss modulus as a function of idle time ($t_i$) for a suspension having 1 mM salt. From right to left: $t_i$ = 6, 12, 15, 18, 27, 52 days. The grey lines superposed on $t_i$ = 18 day dataset are the horizontally shifted evolution curves (Storage and Loss modulus) obtained on other idle times.

We carried out similar superposition procedure for the evolution of $G'$ and $G''$ at various idle times for five salt concentrations in the range 0.1 mM to 7 mM. Figure 5a and 5b shows respective superpositions of $G'$ and $G''$ obtained by shifting all the idle time dependent data on to the data obtained on 18th day ($t_i$=18 days). The same horizontal shift factors are required to obtain superposition of $G'$ as well as $G''$. However, in order to get superposition of $G''$, vertical shifting is needed in addition to the horizontal shifting. Interestingly there is a remarkable similarity in figs. 4 and 5. The former figure represents the evolution of both the moduli at different idle times for 1 mM salt concentration, while fig. 5 represents the evolution of both the moduli for various salt concentrations on the idle time $t_i$=18 days. Interestingly, similar to the behavior observed for the idle time dependence, the evolution of $G'$ and $G''$ also get shifted to the lower age with the increase in the concentration of salt. Additional similarities include two stage increase in $G'$ with transition marked by maximum in $G''$, which also approximately corresponds to crossover of $G'$ and $G''$ (the aging time at which $G''$ shows maximum is approximately 1.5 times larger than the aging time at which $G'$ crosses over $G''$, however these two points appear very close due to the logarithmic nature of the scale). For 7 mM salt concentration system, irrespective of the idle time, the value of $G'$ in the beginning of the experiment is always greater than $G''$ and shows only a second stage of the evolution. Likewise, the corresponding $G''$ also shows only a decrease with the aging time. Importantly, it should be noted that for $C_s$=1 mM and on $t_i$=52 days, the observed evolution of $G'$ and $G''$ shown in fig. 4 is qualitatively similar to the evolution data for 7 mM salt concentration shown in fig. 5.

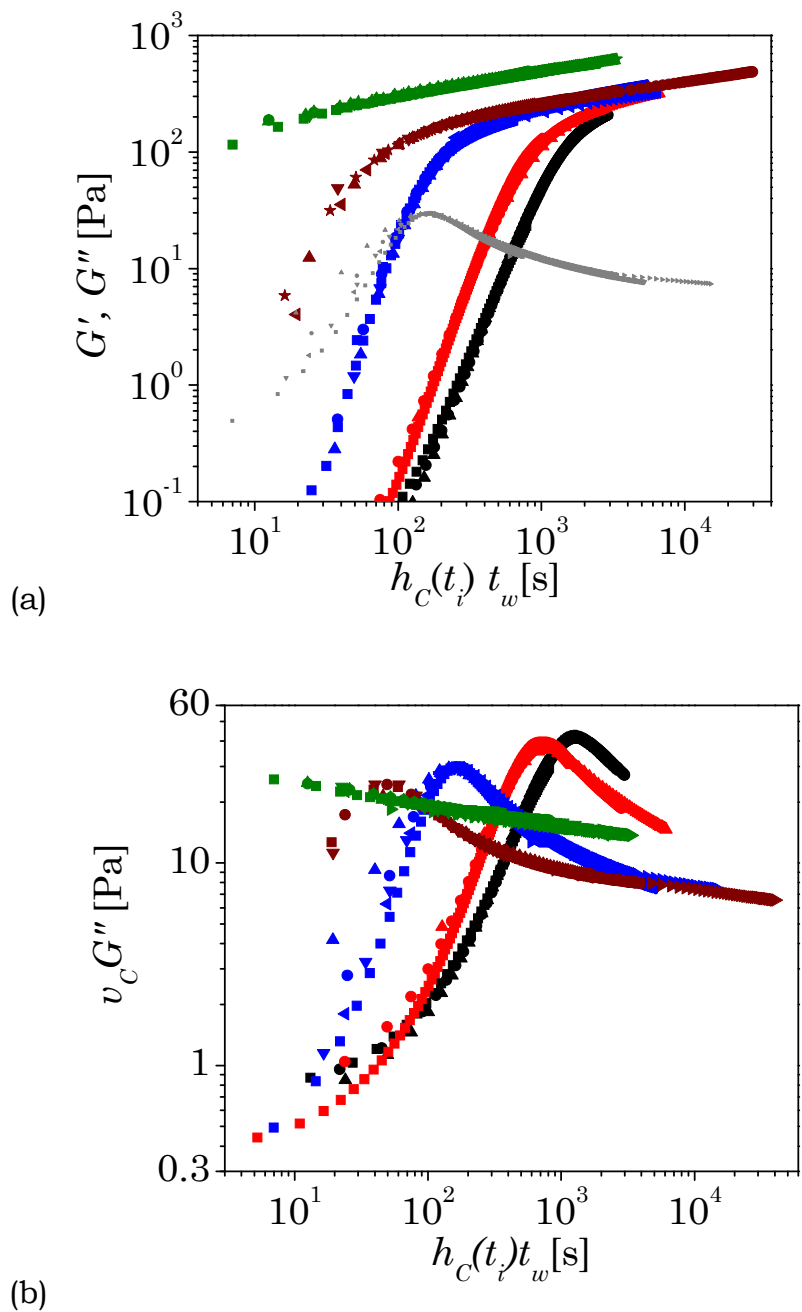

**Figure 5**: Shifted curves of (a) Storage and (b) loss modulus for various salt concentrations. From right to left: 0.1, 1, 3, 5 and 7 mM salt concentration. The same color represents same concentration of salt in both the plots.

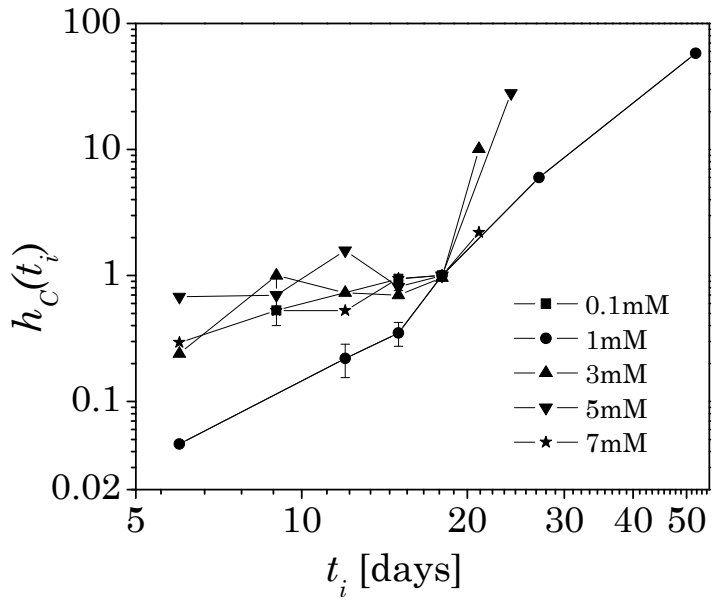

(a)

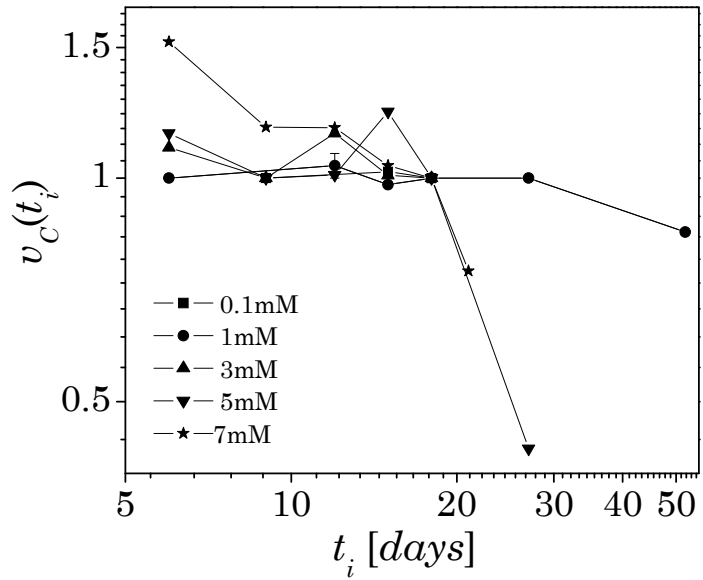

(b)

**Figure 6**: (a) Horizontal shift factor as a function of idle time ($t_i$) for storage and loss modulii. (b) Vertical shift factor as a function of idle time ($t_i$) for loss modulii (Squares: 0.1mM; circles: 1mM; up triangles: 3mM; down triangles 5mM; stars: 7mM).

We have plotted the horizontal and vertical shift factors shown in figs. 5a and 5b as a function of idle time ($t_i$) in figs. 6a and 6b respectively. In general, the horizontal shift factors, which are common to both the moduli, increase with the idle time ($t_i$). This behavior suggests that the evolution of both the moduli at high idle times shifts to the low aging times irrespective of the concentration of salt. The vertical shift factor is needed only for the viscous modulus ($G''$) data to obtain the superposition. As shown in fig. 6b, the vertical shift factors remain almost close to unity for 0.1, 1, and 3 mM systems. However, for 5 and 7 mM salt concentrations, for which most of the evolution is in the non-ergodic regime, the vertical shift factors decrease with the idle time.

We also performed the aging experiments for above mentioned five salt concentrations at various frequencies in the range 0.1 to 10 Hz. It is observed that, for all the explored salt concentrations, evolution of $G'$ becomes independent of frequency after the crossover of $G'$ and $G''$. According to Fielding *et al.*,[53] this observation of weak dependence on frequency after the crossover of $G'$ and $G''$ again confirms the system to be in the non-ergodic state. This weak frequency dependence implies that the shift factors shown in figs. 6a and 6b do not depend on frequency for the part of the superposition in the non-ergodic regime.

The superpositions of the idle time dependent data for the five salt concentration systems, shown in figs. 5a and 5b, demonstrate that the evolutions of both the moduli for the systems having higher concentration of salt get shifted to a lower age preserving the overall curvature. In fig. 7 we have plotted the master superposition by horizontally and vertically shifting the concentration dependent $G'$ and $G''$ superpositions plotted in figs. 5a and 5b. The respective superpositions at various salt concentrations are shifted to the superposition associated with 5 mM system, which results in '*idle time – aging time – salt concentration superposition.*' The corresponding horizontal shift factor, which is common for both the moduli, is plotted as a function of concentration of salt ($C_s$) in fig. 8. It can be seen that the horizontal shift factor increases exponentially with increase in the concentration of salt ($C_s$). This master superposition and the associated trend of the horizontal shift factor implies that the aging that occurs in the low salt concentration systems at greater idle times is equivalent to what occurs in high salt concentration systems at smaller idle times. The vertical shift factors required to obtain the superposition of $G'$ and $G''$ are also plotted in fig. 8. It can be seen that the same vertical shift factor is required to shift both $G'$ and $G''$ data except that for 7 mM salt concentration system. The greater values of $G''$ evolution for $C_s$ =7 mM system compared to other salt concentrations, as shown in fig. 4b., lead to the vertical shift factor for $G''$ data for $C_s$ =7 mM system to be much less than the overall trend. Nevertheless, other than this exception, the vertical shift factors increase with increase in $C_s$.

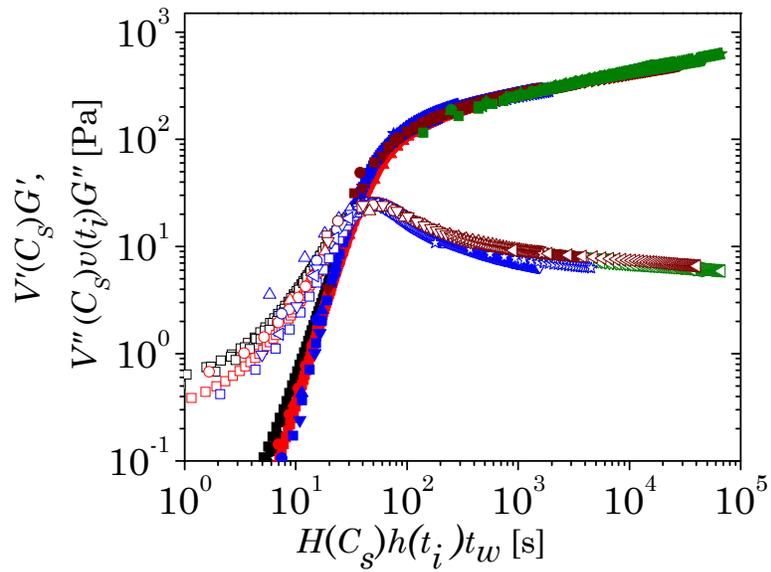

**Figure 7:** A complete master curve combining both idle time ($t_i$) and salt concentration ($C_s$) dependence for representing evolution of (a) storage (filled symbols) and (b) loss modulus.(open symbols)

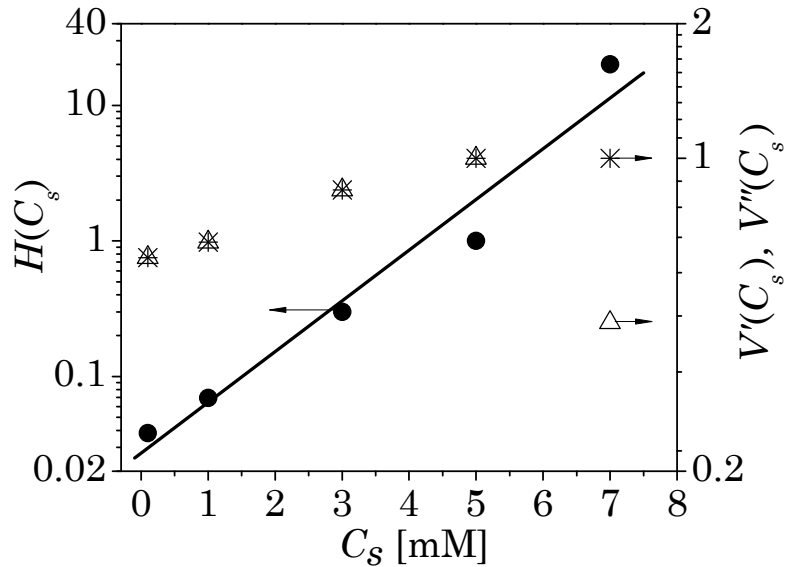

**Figure 8:** Horizontal and vertical shift factors used for obtaining the overall master curve for various salt concentration. Horizontal shift factor, $H$ (filled circles), which is common to both $G'$ and $G''$, shows exponential dependence on salt concentration. The vertical shift factor $V'$ is represented by stars while $V''$ is represented by open triangles.

**IV. Discussion**

In this section we will discuss the implications of the experimental results presented in the previous section with respect to the microstructure and the aging behavior of aqueous suspension of Laponite. As mentioned before, the microstructure of the aqueous suspension of Laponite is influenced by its anisotropic shape, uneven charge distribution on the particle, pH of the medium, and the concentration of dissociated $Na^+$ ions (either counterions or ions due to externally dissolved NaCl). These variables lead to the possibility of three types of inter-particle interactions among the Laponite particles. The first interaction is the electrostatic repulsion between the negatively charged surfaces due to overlapping double layers.[11, 20-22, 42, 49] The second proposed interaction is the attraction between the negatively charged surface and the positively charged edge.[18, 36, 44] The third proposed interaction is the van der Waals attraction between the Laponite discs at higher concentrations of salts when the negative charge on the Laponite surface is shielded by cations.[36, 54] In addition, the sample preparation protocol also has independent effect on the nature of the arrested state.[12, 18] Although the suspensions prepared using any protocol will involve mainly the above mentioned inter-particle interactions; one protocol may lead the suspension to a certain section of the energy landscape, which may not be accessible through another protocol.

It is usually believed in the literature that for the systems with low concentration of cations, overall interactions among the Laponite particles is primarily repulsive.[13, 14, 25, 30, 42, 49, 54-56] Increase in the concentration of cations (addition of salt) tends to shield the repulsive interactions thereby reducing the energy barrier for the particles to approach each other. In order to get a quantitative feel for the repulsive interactions among the Laponite particles, we estimated the Debye-Hückel screening length (characteristic length of the double layer, $1/\kappa$) of the same for the aqueous suspension without any externally added salt. For this system, the suspension is expected to contain $Na^+$ ions dissociated from faces and $OH^-$ ions dissociated from the edges. If we assume that the major contributions to the ionic conductivity ($\sigma$) is from concentration of $Na^+$ ions ($\sigma \approx \sigma_{Na}$), concentration of sodium ions $n_{Na}$ can be estimated as: $n_{Na} = \sigma_{Na}/\mu_{Na}e$, where $\mu_{Na}$ is mobility of $Na^+$ ions and $e$ is the electron charge.[15] The conductivity of 2.8 wt % Laponite suspension having no externally added salt on 14th day is 848μS/cm, which leads to Debye screening length of 3.22 nm. We also measured conductivity values for other salt concentration systems, which are 904, 1119, 1258, 1576μS/cm for suspension containing 1, 3, 5 and 7 mM NaCl respectively. Aqueous suspension having NaCl, contains $Na^+$ as well as $Cl^-$ ions and since both the ions contribute to the conductivity, it is not straightforward to estimate the number density of $Na^+$ ions (concentration of $Cl^-$ ions = concentration of NaCl). Nonetheless, it is apparent that the Debye screening length of the systems with higher concentration of salt should to be much smaller than the same for no salt system (3.22 nm). It must be mentioned that, in the literature, various groups have used different values of the number density of $Na^+$ ions to estimate the Debye screening length. Some groups use the value only based on the externally added salt while some groups use the actual value which also includes the counterion concentration. We have employed the later approach in this work, which is also consistent with number of previous studies.[15, 18, 36, 57] Therefore, this work suggests that, even for the suspension without salt, the length-scale associated with the repulsive interactions (3.22 nm) is much smaller than the diameter of the particle (30 nm).

The Laponite particle in low salt concentration aqueous suspension is considered to be trapped in a cage formed by the neighboring particles. In an aging process every trapped particle undergoes a microscopic motion of structural rearrangement which takes it to a lower potential energy state.[1] Such exploration of the lower energy states as a function of time manifests in gradual increase in the elastic modulus, which is a typical characteristic feature of the aging process in soft materials.[6, 14, 58] Our observation of higher shear melting viscosity for the experiments carried out at greater idle times suggests that the low energy state is associated with the formation of strong interactions among the Laponite particles, which are difficult to obliterate by high shear stress applied in the shear melting step. Similar to the idle time dependence, evolution of the suspension having higher salt concentration also gets shifted to lower aging times. Interestingly both the idle time dependence and salt concentration dependence demonstrate a superposition, suggesting that the low energy state associated with aqueous suspension of Laponite is kinetically affected by the concentration of salt. Interestingly the exponential dependence of the horizontal shift factor on the salt concentration $\left[\ln H(C_s) \sim C_s\right]$, shown in fig. 8 suggests that the barrier height to acquire a low energy state decreases linearly with the salt concentration.[59] Interestingly Mongondry and coworkers[18] used dynamic light scattering and showed that the time required to reach the a certain level of reduced scattering intensity decreases exponentially with increase in the concentration of salt demonstrating an exponential dependence of the kinetics on the salt concentration $\left[t \sim 10^{-0.42 C_s}\right]$. Remarkably their coefficient 0.42 matches closely with the coefficient 0.37 observed in present study (figure 8).

Laponite, due to its oblate shape is expected to show a nematic order in the aqueous suspensions at higher concentrations. Lemaire et al.[60] and Gabriel et al.[4] observed an orientational order in aqueous suspension of Laponite B above 2 weight % concentration. It should be noted that unlike Laponite RD, a synthetic layered silicate, used in this work; Laponite B is a synthetic layered fluorosilicate and has diameter 40 nm. Isotropic to biphasic and biphasic to nematic transition concentrations in oblate systems with hard surface interactions can be estimated by using Onsager's approach.[61] If $a$ is the aspect ratio ($L/D$), the volume fractions at which discotic system undergoes isotropic to biphasic and biphasic to nematic transition are given by $\phi_{i-b}=0.33\,a$ and $\phi_{b-n}=0.45\,a$ respectively.[61] For Laponite B, $a=1/40$ leads to $\phi_{i-b}=0.00825$, which gives weight fraction ≈ 0.02. Remarkably this is precisely the same weight fraction for which both the above mentioned papers observe order in their Laponite B suspension. Lamaire et al.[60] observed the nematic order parameter of 0.55±0.05 for 3 wt. % Laponite B suspension. In principle the structure factor should be above 0.8 for such transition in hard disks.[61] They attribute this low order parameter to the topological defects that remain in their oriented samples.[60] For Laponite RD, various experimental studies have proposed 3 to 3.5 wt. % to be the concentration range above which order exists in the Laponite particles.[20, 27] In order to investigate state of Laponite suspension studied in the present work, we studied the birefringence behavior of the aged samples by placing the same between two crossed polarizers. Interestingly we observed birefringence in all the explored salt concentration systems suggesting our aged 2.8 weight % system to have an orientational order. If we consider Laponite RD to have diameter of 30 nm (usually between 25 to 30 nm), Onsager's approach gives weight fraction above which order should be observed to be 0.0275 (2.75 weight %). However in case of Laponite, an edge of the particle is positively charged while the surface has a negative charge. In addition the system is out of equilibrium and physically jammed. Under such situations the Onsager's approach which is strictly based on hard surface interactions and proposes liquid – to – liquid crystalline nematic first order thermodynamic transition may not be applicable. Nonetheless the close match of the concentrations at which the order is observed in Laponite B as well as Laponite RD with that of the scaling arguments of Onsager is striking.

In the literature various possible structural scenarios for the aqueous suspension of Laponite have been proposed as a function of concentration of Laponite and concentration of NaCl. Since the edge of a Laponite particle is positively charged while the face has a permanent negative charge, one of the proposals suggests that aging in this system involves the Laponite particles reorienting themselves slowly to form the edge-to-surface interactions as a function of time.[18, 36, 44] It is believed that the positive edge – negative surface interaction between the Laponite particles is significantly strong.[18] Therefore, there is a possibility that while rejuvenating the overall suspension, the high stress cannot break all the edge-to-surface interactions causing partial irreversibility. With increase in the idle time, number density of such interactions increases enhancing the number of edge-surface interactions that cannot be broken by the high stress, causing a steady enhancement in the plateau viscosity of the suspension in the shear melting step. In the subsequent aging experiments, since the aging process begins at a more matured (low energy) state; for the experiments carried out at a greater idle time, the overall evolution gets shifted to a lower age. As mentioned before, increase in the concentration of NaCl reduces the repulsion among the Laponite particles. Therefore, according to edge-to-surface aggregation school (house of cards structure), with the decrease in the extent of repulsion among Laponite particles, the rate of formation of edge-to-surface interactions is also expected to increase. This should lead to a faster build up of the structure for the experiments carried out at a higher concentration of salt making the evolution of both the moduli shift to a lower age. At very high concentration of salt, the formation of the edge-to-surface interactions is so fast that system directly enters the nonergodic state $(G' > G'')$ after the shear melting stops.

Therefore, even though the time taken by the low salt concentration suspension to reach a certain modulus level is significantly large compared to the same for high salt concentration system, the overall evolution maintains the same curvature on a logarithmic time axis leading to the superposition shown in fig. 7, thereby explaining the observed rheological behavior.

While the house of cards proposal (edge-to-surface interactions) explains the observed irreversibility in the aging behavior in the rheological experiments carried out in this work, the observed anisotropy in the Laponite suspension poses a strong challenge to this proposal. In principle the collapsed house of cards structure (where the contact between two Laponite disks is bent) may explain the observed order and lowering of the nematic order parameter. However, in order to form such structure with bent contacts that fills the whole space, enough number density of Laponite particles is needed in the system. If we consider $n$ to be number of Laponite particles per unit volume, the volume fraction ($\phi_v$) of Laponite in water, in the limit of low volume fraction of Laponite, is given by: $\phi_v = n\pi R^2 h$, where $R$ is radius of Laponite disc (=13 to 15 nm) and $h$ is thickness of Laponite particle (~1 nm). Furthermore, the volume fraction of the spheres encompassing the $n$ particles is given by: $\frac{4}{3}n\pi R^3 = 4\phi_v R/(3h)$. For a system having concentration of 2.8 weight % the volume fraction of Laponite in water is around 0.011, which leads to the volume fraction of the spheres encompassing the particles at this concentration to 0.22. This volume fraction is too low to form space filling structure with bent interparticle contacts necessary to demonstrate an order. Moreover, the average interparticle distance for disk like particles is given by: $\left(\pi R^2 h/\phi_v\right)^{1/3}$.[15] This leads to an average inter-particle distance of around 40 nm for a system having concentration of 2.8 weight %, which is greater than the diameter of the particle. Therefore, this analysis suggests collapsed house of cards structure with bent angle of contact may not form a space filling arrested system at the concentration of 2.8 weight %.

To summarize discussion in this section, it is apparent that the length-scale associated with repulsive interactions among the Laponite particles due to electrostatic screening is much smaller than the diameter of the particle for all the studied concentrations of salt. In addition the edge has positive charge. The possible house of cards structure where there is a bond between positive edge and negative surface explains irreversible aging and the salt concentration dependence observed in the rheological experiments, but does not explain the origin of anisotropy in the structure. Therefore, the possible microstructure of the system should be such that the particles remain in a long range orientational order with inter-particle interactions, source of which is not yet clear to us, causing non-ergodicity. The corresponding exploration of an energy landscape with respect to aging time by the system then occurs in such unique way that does not allow reversibility upon application of deformation field and demonstrates the kinetic dependence on the salt concentration. We feel that in order to get more insights into the present system the rheological and birefringence experiments performed in this work are needed to be carried out over a broader spectrum of concentrations of Laponite. In addition, there is clearly a need for more work on this topic particularly on realistic simulations to propose such microstructures as a function of system variables that will unify various observations of scattering experiments and rheological experiments mentioned in the literature.

**V. Conclusions**

In this work we study the aging behavior of aqueous suspension of Laponite having 2.8 wt. % concentration as a function of the idle time (time elapsed since preparation of the suspension) and the concentration of NaCl. Interestingly aged samples of Laponite suspension at all the studied salt concentrations show birefringence when observed using crossed polarizers suggesting anisotropic nature of the microstructure of the same. In rheological experiments, we observe that the plateau value of the complex viscosity obtained in the shear melting experiments carried out at greater idle times increases for all the explored salt concentrations, thereby demonstrating partial irreversibility in aging. In addition, the subsequent evolution of the elastic and the viscous modulus, which represents aging, shifts to lower aging times for the experiments carried out at greater idle times. Interestingly all the idle time dependent evolution data shows the superposition suggesting time separated self similar aging behavior. In the salt concentration dependent experiments, evolution of elastic and viscous moduli shifts to lower aging times for higher concentration of salt thereby showing kinetic dependence of rate of aging on the salt concentration. Remarkably horizontal shifting of the salt concentration and idle time dependent data shows the *'idle time – aging time – salt concentration superposition'* suggesting self similarity of the aging dynamics in aqueous suspension of Laponite. We estimate that the energy barrier to explore the low energy states during aging process decreases linearly with increase in the concentration of salt. Remarkably the idle time and the salt concentration dependent evolution of both the moduli superpose to show a generic trend. These results imply that the aging that occurs over a very long period in low salt concentration systems is qualitatively similar to that occurs in high salt concentration systems over a short period in the aqueous suspension of Laponite.

**Acknowledgement**: Financial support from Department of Science and Technology, Government of India, under IRHPA scheme is greatly acknowledged. We thank Prof. G. Anantharaman, Prof. A. Ramakrishnan, Dr. Guruswamy for the discussion. We also thank department of Chemistry, IIT Kanpur for providing the facility to carry out complexometric titration and Prof. Asima Pradhan for making available the cross polarizer setup.

**References:**
1. Wales, D. J., *Energy Landscapes*. Cambridge University Press: Cambridge, 2003.
2. Van Olphen, H., *An Introduction to Clay Colloid Chemistry*. Wiley: New York, 1977.
3. Meunier, A., *Clays*. Springer: Berlin, 2005.

4. Gabriel, J.-C. P.; Sanchez, C.; Davidson, P., Observation of nematic liquid-crystal textures in aqueous gels of smectite clays. *J. Phys. Chem.* **1996,** 100, (26), 11139-11143.
5. Shalkevich, A.; Stradner, A.; Bhat, S. K.; Muller, F.; Schurtenberger, P., Cluster, Glass, and Gel Formation and Viscoelastic Phase Separation in Aqueous Clay Suspensions. *Langmuir* **2007,** 23, (7), 3570-3580.
6. Awasthi, V.; Joshi, Y. M., Effect of temperature on aging and time-temperature superposition in nonergodic laponite suspensions. *Soft Matter* **2009,** 5, (24), 4991-4996.
7. Katzel, U.; Richter, T.; Stintz, M.; Barthel, H.; Gottschalk-Gaudig, T., Phase transitions of pyrogenic silica suspensions: A comparison to model laponite. *Physical Review E - Statistical, Nonlinear, and Soft Matter Physics* **2007,** 76, (3).
8. Li, L.; Harnau, L.; Rosenfeldt, S.; Ballauff, M., Effective interaction of charged platelets in aqueous solution: Investigations of colloid laponite suspensions by static light scattering and small-angle x-ray scattering. *Phys. Rev. E* **2005,** 72, (5), 051504.
9. Odriozola, G.; Romero-Bastida, M.; Guevara-Rodrìguez, F. D. J., Brownian dynamics simulations of Laponite colloid suspensions. *Physical Review E - Statistical, Nonlinear, and Soft Matter Physics* **2004,** 70, (2 1), 021405-1-021405-15.
10. Baghdadi, H. A.; Parrella, J.; Bhatia, S. R., Long-term aging effects on the rheology of neat laponite and laponite - PEO dispersions. *Rheologica Acta* **2008,** 47, (3), 349-357.
11. Bonn, D.; Kellay, H.; Tanaka, H.; Wegdam, G.; Meunier, J., Laponite: What is the difference between a gel and a glass? *Langmuir* **1999,** 15, (22), 7534-7536.
12. Cummins, H. Z., Liquid, glass, gel: The phases of colloidal Laponite. *Journal of Non-Crystalline Solids* **2007,** 353, (41-43), 3891-3905.
13. Joshi, Y. M., Model for cage formation in colloidal suspension of laponite. *The Journal of Chemical Physics* **2007,** 127, (8), 081102.
14. Joshi, Y. M.; Reddy, G. R. K.; Kulkarni, A. L.; Kumar, N.; Chhabra, R. P., Rheological Behavior of Aqueous Suspensions of Laponite: New Insights into the Ageing Phenomena. *Proc. Roy. Soc. A* **2008,** 464, 469-489.
15. Jabbari-Farouji, S.; Tanaka, H.; Wegdam, G. H.; Bonn, D., Multiple nonergodic disordered states in Laponite suspensions: A phase diagram. *Phys. Rev. E* **2008,** 78, (6), 061405-10.
16. Knaebel, A.; Bellour, M.; Munch, J.-P.; Viasnoff, V.; Lequeux, F.; Harden, J. L., Aging behavior of Laponite clay particle suspensions. *Europhysics Letters* **2000,** 52, (1), 73-79.
17. Levitz, P.; Lecolier, E.; Mourchid, A.; Delville, A.; Lyonnard, S., Liquid-solid transition of Laponite suspensions at very low ionic strength: Long-range electrostatic stabilisation of anisotropic colloids. *Europhysics Letters* **2000,** 49, (5), 672-677.
18. Mongondry, P.; Tassin, J. F.; Nicolai, T., Revised state diagram of Laponite dispersions. *J.Colloid Interface Sci.* **2005,** 283, (2), 397-405.
19. Mossa, S.; De Michele, C.; Sciortino, F., Aging in a Laponite colloidal suspension: A Brownian dynamics simulation study. *Journal of Chemical Physics* **2007,** 126, (1), 014905.
20. Mourchid, A.; Delville, A.; Lambard, J.; Lecolier, E.; Levitz, P., Phase diagram of colloidal dispersions of anisotropic charged particles: Equilibrium properties, structure, and rheology of laponite suspensions. *Langmuir* **1995,** 11, (6), 1942-1950.
21. Mourchid, A.; Delville, A.; Levitz, P., Sol-gel transition of colloidal suspensions of anisotropic particles of laponite. *Faraday Discuss.* **1995,** 101, 275-285.
22. Mourchid, A.; Lecolier, E.; Van Damme, H.; Levitz, P., On viscoelastic, birefringent, and swelling properties of laponite clay suspensions: Revisited phase diagram. *Langmuir* **1998,** 14, 4718-4723.
23. Mourchid, A.; Levitz, P., Long-term gelation of laponite aqueous dispersions. *Phys. Rev. E* **1998,** 57, R4887-R4890.
24. Nicolai, T.; Cocard, S., Light scattering study of the dispersion of laponite. *Langmuir* **2000,** 16, (21), 8189-8193.
25. Reddy, G. R. K.; Joshi, Y. M., Aging under stress and mechanical fragility of soft solids of laponite. *Journal of Applied Physics* **2008,** 104, 094901.
26. Ruzicka, B.; Zulian, L.; Ruocco, G., Ergodic to non-ergodic transition in low concentration, Laponite. *J. Phys. Con. Mat.* **2004,** 16, (42), S4993.
27. Ruzicka, B.; Zulian, L.; Ruocco, G., More on the phase diagram of laponite. *Langmuir* **2006,** 22, (3), 1106-1111.
28. Ruzicka, B.; Zulian, L.; Ruocco, G., Ageing dynamics in Laponite dispersions at various salt concentrations. *Philosophical Magazine* **2007,** 87, (3-5), 449-458.
29. Saunders, J. M.; Goodwin, J. W.; Richardson, R. M.; Vincent, B., A Small-Angle X-ray Scattering Study of the Structure of Aqueous Laponite Dispersions. *The Journal of Physical Chemistry B* **1999,** 103, (43), 9211-9218.
30. Schosseler, F.; Kaloun, S.; Skouri, M.; Munch, J. P., Diagram of the aging dynamics in laponite suspensions at low ionic strength. *Phys. Rev. E* **2006,** 73, 021401.
31. Zulian, L.; Ruzicka, B.; Ruocco, G., Influence of an adsorbing polymer in the aging dynamics of Laponite clay suspensions. *Philosophical Magazine* **2008,** 88, 4213-4221.
32. Joshi, Y. M., Modeling Dependence of Creep Recovery Behavior on Relaxation Time Distribution of Aging Colloidal Suspensions. *Industrial & Engineering Chemistry Research* **2009,** 48, (17), 8232-8236.
33. Labanda, J.; Llorens, J., Effect of aging time on the rheology of Laponite dispersions. *Colloids and Surfaces A: Physicochemical and Engineering Aspects* **2008,** 329, (1-2), 1-6.
34. www.laponite.com
35. Kroon, M.; Vos, W. L.; Wegdam, G. H., Structure and formation of a gel of colloidal disks. *Phys. Rev. E* **1998,** 57, 1962-1970.
36. Tawari, S. L.; Koch, D. L.; Cohen, C., Electrical double-layer effects on the Brownian diffusivity and aggregation rate of Laponite clay particles. *Journal of Colloid and Interface Science* **2001,** 240, (1), 54-66.
37. Tombácz, E.; Szekeres, M., Colloidal behavior of aqueous montmorillonite suspensions: the specific role of pH in the presence of indifferent electrolytes. *Applied Clay Science* **2004,** 27, (1-2), 75-94.
38. Kosmulski, M., *Chemical Properties of Material Surfaces* Marcel Dekker: New York, 2001.
39. Martin, C.; Pignon, F.; Piau, J.-M.; Magnin, A.; Lindner, P.; Cabane, B., Dissociation of thixotropic clay gels. *Phys. Rev. E* **2002,** 66, (2), 021401.
40. *Laponite Technical Bulletin: Laponite: Structure, chemistry and relationship to natural clays.* 1990; Vol. L104-90-A, p 1-15.

41. Thompson, D. W.; Butterworth, J. T., The nature of laponite and its aqueous dispersions. *J. Colloid Interface Sci.* **1992,** 151, (1), 236-243.
42. Tanaka, H.; Meunier, J.; Bonn, D., Nonergodic states of charged colloidal suspensions: Repulsive and attractive glasses and gels. *Phys. Rev. E* **2004,** 69, (3 1), 031404.
43. Bonn, D.; Tanaka, H.; Wegdam, G.; Kellay, H.; Meunier, J., Aging of a colloidal "Wigner" glass. *Europhys. Lett.* **1998,** 45, 52-57.
44. Dijkstra, M.; Hansen, J.-P.; Madden, P. A., Statistical model for the structure and gelation of smectite clay suspensions. In *Phys. Rev. E*, 1997; Vol. 55, pp 3044-3053.
45. Jabbari-Farouji, S.; Wegdam, G. H.; Bonn, D., Gels and Glasses in a Single System: Evidence for an Intricate Free-Energy Landscape of Glassy Materials. *Phys. Rev. Lett.* **2007,** 99, (6), 065701-4.
46. Shukla, A.; Joshi, Y. M., Ageing under oscillatory stress: Role of energy barrier distribution in thixotropic materials. *Chemical Engineering Science* **2009,** 64, (22), 4668-4674.
47. Cocard, S.; Tassin, J. F.; Nicolai, T., Dynamical mechanical properties of gelling colloidal disks. *J. Rheol.* **2000,** 44, (3), 585-594.
48. Bhatia, S.; Barker, J.; Mourchid, A., Scattering of disklike particle suspensions: Evidence for repulsive interactions and large length scale structure from static light scattering and ultra-small-angle neutron scattering. *Langmuir* **2003,** 19, (3), 532-535.
49. Bonn, D.; Tanaka, H.; Wegdam, G.; Kellay, H.; Meunier, J., Aging of a colloidal "Wigner" glass. *Europhysics Letters* **1999,** 45, (1), 52-57.
50. Ruzicka, B.; Zulian, L.; Angelini, R.; Sztucki, M.; Moussaid, A.; Ruocco, G., Arrested state of clay-water suspensions: Gel or glass? *Phys. Rev. E* **2008,** 77, (2), 020402-4.
51. Vogel, A. I., *A textbook of quantitative inorganic analysis*. Longman: New York, 1978.
52. Ovarlez, G.; Coussot, P., Physical age of soft-jammed systems. *Phys. Rev. E* **2007,** 76, (1), 011406.
53. Fielding, S. M.; Sollich, P.; Cates, M. E., Aging and rheology in soft materials. *J. Rheol.* **2000,** 44, (2), 323-369.
54. Tanaka, H.; Jabbari-Farouji, S.; Meunier, J.; Bonn, D., Kinetics of ergodic-to-nonergodic transitions in charged colloidal suspensions: Aging and gelation. *Phys. Rev. E* **2005,** 71, (2), 021402.
55. Bellour, M.; Knaebel, A.; Harden, J. L.; Lequeux, F.; Munch, J.-P., Aging processes and scale dependence in soft glassy colloidal suspensions. *Phys. Rev. E* **2003,** 67, 031405.
56. Kaloun, S.; Skouri, R.; Skouri, M.; Munch, J. P.; Schosseler, F., Successive exponential and full aging regimes evidenced by tracer diffusion in a colloidal glass. *Phys. Rev. E* **2005,** 72, 011403.
57. Bosse, J.; Wilke, S. D., Low-density ionic glass. *Phys. Rev. Lett.* **1998,** 80, (6), 1260-1263.
58. Cloitre, M.; Borrega, R.; Leibler, L., Rheological aging and rejuvenation in microgel pastes. *Phys. Rev. Lett.* **2000,** 85, (22), 4819-4822.

59. The horizontal shift factor is expected to depend on the barrier height ($U_B$) as: $H \sim \exp[-U_B/k_B T]$. The present relation: $(\ln[H(C_s)] \sim C_s)$; therefore suggests that the barrier height decreases linearly with $C_s$.
60. Lemaire, B. J.; Panine, P.; Gabriel, J. C. P.; Davidson, P., The measurement by SAXS of the nematic order parameter of laponite gels. *Europhysics Letters* **2002,** 59, (1), 55-61.
61. De Gennes, P. G.; Prost, J., *The Physics of Liquid Crystals*. Oxford: New York, 1993.